\documentstyle[erice98,psfig,12pt]{article}
\begin{document}

\title{Reaction dynamics in Pb+Pb at the CERN/SPS:\\ from partonic degrees
	of freedom to freeze-out
 } 

\author{S.A. Bass\footnote{Feodor Lynen Fellow of the Alexander v. Humboldt
	Foundation}}
\address{
	Department of Physics\\
	Duke University\\
	Durham, NC, 27708-0305, USA
	}

\author{H.~Weber, C.~Ernst, M.~Bleicher, 
	M.~Belkacem$^2$, L.~Bravina\footnote{Alexander v. Humboldt Fellow},
	S.~Soff, H.~St\"ocker and W.~Greiner}
\address{
  	Institut f\"ur Theoretische Physik \\
	Johann Wolfgang Goethe Universit\"at\\
	Robert Mayer Str. 8-10\\
	D-60054 Frankfurt am Main, Germany}

\author{C. Spieles$^1$}
\address{
	Lawrence Berkeley Laboratory\\
	1 Cyclotron Road\\
	Berkeley, CA 94720, USA
	}

\maketitle

\begin{abstract}
We analyze the reaction dynamics of central Pb+Pb collisions
at 160 GeV/nucleon. First
we estimate the energy density $\epsilon$ pile-up at mid-rapidity
and calculate its excitation function: 
$\epsilon$ is decomposed into hadronic and partonic contributions.
A detailed analysis of the collision dynamics in the framework of a
microscopic transport model shows the importance of partonic degrees
of freedom and rescattering of leading (di)quarks in the early 
phase of the reaction for $E_{lab} \ge 30$~GeV/nucleon. 
The energy density reaches up to
4 GeV/fm$^3$, 95\% of which are contained in partonic degrees of freedom.
It is shown that cells of hadronic matter, after $t \approx 2 R/\gamma v_{cm}$,
can be viewed as nearly chemically equilibrated. This matter never
exceeds energy densities of $\sim 0.4$~GeV/fm$^{3}$, i.e. a density
above which the notion of separated hadrons loses its meaning. 
The final reaction stage is analyzed in terms of hadron ratios, 
freeze-out distributions and a source analysis for final state pions.
\end{abstract}

\newpage

The study of relativistic heavy ion collisions offers the unique 
opportunity to study hot and dense QCD matter under conditions which
are thought to have existed in the early stages of our universe.
However, only the hadronic final state of the heavy ion collision 
is accessible via experiment, or -- in the case of leptonic probes --
the time integral of the emission over the entire reaction history.
Microscopic transport models offer the unique opportunity to link
this final state information to the experimentally inaccessible 
early and intermediate reaction stages. In this paper we analyze
the reaction dynamics of central Pb+Pb collisions at CERN/SPS energies.
We focus specifically on the time evolution of energy density and
its interpretation in terms of hadronic and partonic degrees of freedom.
We then discuss the possible formation of a thermally and chemically
equilibrated state in the central reaction zone and finally investigate
the late reaction stages with a decomposition of freeze-out radii
and sources for individual hadron species.

The determination of energy densities in ultra-relativistic heavy-ion
collisions is crucial for any discussion involving a possible deconfinement
phase transition to a QGP 
\cite{collins75a,stoecker80a,shuryak80a,mclerran86a,stoecker86a,clare86a,kajantie87a}.
Estimates for the energy density
during the hot and dense early reaction stage have been given by a large
variety of different models 
\cite{bjorken83a,stoecker86a,clare86a,vanhove82a,mclerran86a,bleicher98a}.

It has been questioned whether hadronic transport models are still valid
at CERN/SPS energies: the energy density estimates obtained in these 
frameworks are claimed to be well above the critical energy density
estimates for a deconfinement phase transition given by 
Lattice Gauge Theory \cite{blum95a,boyd95a,laermann96a}.
Hadronic transport models, however, contain implicit partonic
degrees of freedom -- particle production at high energies is e.g. 
modeled via the production and fragmentation of strings
\cite{andersson87a,andersson87b,sjoestrand94a}. 

In the UrQMD model used below, the leading
hadrons of the fragmenting string contain the valence-quarks of the
original excited hadron. These leading hadrons are allowed -- in the model -- 
to interact even during their formation time, with a reduced cross section, 
thus accounting for the original valence quarks contained in that
hadron. Those leading hadrons represent a simplified
picture of the leading (di)quarks of the fragmenting string. 
Newly to-be-produced hadrons which do not contain string valence quarks
do in the present model not interact during their formation time -- however,
they contribute to the energy density of the system.
A proper treatment of the partonic degrees of freedom during the
formation time ought to include soft and hard parton scattering
\cite{geiger92a} and the explicit time-dependence of the color interaction
between the expanding quantum wave-packets \cite{gerland98a}:
However, such an improved treatment of the internal hadron dynamics has
not been implemented for light quarks into the present model.
Therefore, in the following analysis all contributions stemming from
hadrons within their formation time are termed ``partonic''. 
All contributions stemming from fully formed hadrons are termed
``hadronic''. The main focus of this paper is on the partitioning
and  the time evolution
of the energy density and the collision
dynamics of the early, intermediate, and late reaction stage at 
energies $E_{lab} = 10 - 200$~GeV/nucleon.

The UrQMD model \cite{bass98a} is
based on analogous principles as 
(Relativistic) Quantum Molecular Dynamics
\cite{peilert88a,hartnack89b,aichelin91a,sorge89a,lehmann95a}.
Hadrons are represented by Gaussians   
in phase space and are propagated according
to Hamilton's equation of motion.
The collision term of the UrQMD model treats 55 different
isospin (T) degenerate baryon (B) species
(including nucleon-, delta- and hyperon- resonances with masses up to 2 GeV)
and 32 different T-degenerate meson (M) species,
including (strange) resonances as well as their
corresponding anti-particles, i.e.
full baryon-antibaryon symmetry is included.
Isospin is treated explicitly.
For hadronic excitations with masses $m>2$ GeV (B) and $>1.5$ GeV (M)
a string model is used. Particles produced in the string fragmentation
are assigned a formation time. 
This time $\tau_f$ physically consists of a quantal time $\tau_Q$, i.e.
before the partons are produced, $\tau_Q \sim 1/m$, and a quantum
diffusion time, $\tau_D$, during which the partons evolve in the
medium to build up their internal asymptotic wave-functions to form
the hadron. $\tau_Q$ and $\tau_D$ differ for different parton and hadron
species. For our present purpose, we -- for the sake of simplicity --
just collect all partons, formed and unformed, as one species.
For a detailed overview of the elementary cross sections and string excitation
scheme included in the UrQMD model, see ref. \cite{bass98a}.

\begin{figure}[tb]
\centerline{\psfig{figure=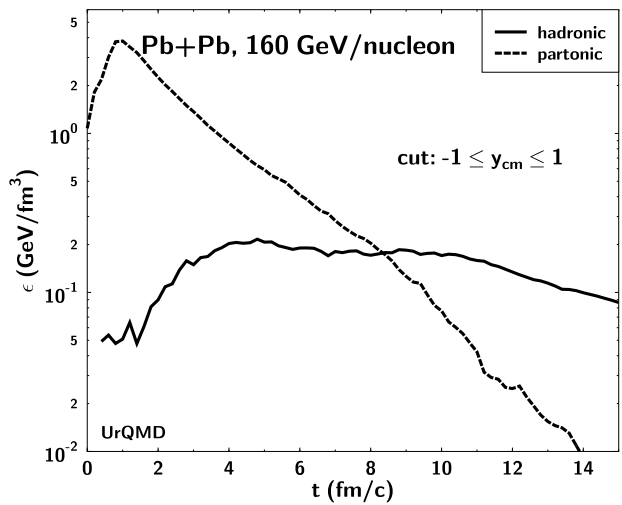,width=3.5in}}
\caption{\label{edens} Time evolution of the energy density $\epsilon$
	in central Pb+Pb reactions at 160 GeV/nucleon. $\epsilon$ has
	been decomposed into ``partonic'' and ``hadronic'' contributions
	and only particles around mid-rapidity have been taken into account.
	The early and intermediate reaction stages are dominated by
	the ``partonic'' contribution.}
\end{figure}

The partitioning of the distinct constituents can be inspected in 
Figure \ref{edens} which shows the time-evolution of the energy 
density for central
Pb+Pb collisions at 160 GeV/nucleon. The nuclei are initialized such that
they touch a $t=0$ fm/c.
The energy density is partitioned into 
the above defined ``hadronic'' contribution, 
from fully formed hadrons, and the ``partonic'' contribution, 
from partons, constituent quarks and diquarks 
within the hadron formation time. 
Nearly all incident baryons are rapidly excited
into strings.  Subsequently,  ``partonic'' energy density builds up,
reaching values
of 4 GeV/fm$^3$ around midrapidity, $\Delta y = 1$ (at $t\approx 1$ fm/c). 
In the course of the reaction
hadrons are formed which increases in the ``hadronic''
energy density, accompanied by a nearly exponential decrease 
in the ``partonic'' energy density.

These energy densities are calculated as follows: In the UrQMD model hadrons
are represented by Gaussian wave packets. The width of the Gaussians
$\sigma=1.04$ fm and their normalization are chosen such that a calculation of 
the baryon density in the initial nuclei  yields ground state nuclear
matter density.
The (energy-) densities in the central reaction zone 
are obtained by summing analytically over all 
Gaussian  hadrons around mid-rapidity
($y_{c.m.} \pm 1$) at the locations of 
these hadrons and then averaging over these energy densities.
This summation over Gaussians  yields a smooth estimate
for baryon- and energy-densities, as compared to counting hadrons in a 
test volume. The rapidity cut insures that only those particles are taken
into account which have interacted. 
Thus, the  free streaming
``spectator'' matter is discarded.

The absolute value of the energy density, however, may depend on the
rapidity cut: Without rapidity cut the energy densities during the
early reaction stage ($t \approx 1$ fm/c) can be as high as 20 GeV/fm$^3$.
Even higher values in $\epsilon$ can
be obtained by choosing the geometric center of the collision
for the sum over the Gaussians instead of averaging over the
energy densities at the locations of the hadrons. The energy
density at a single point may not be physically meaningful
and therefore the latter method is favorable. 

\begin{figure}[tb]
\centerline{\psfig{figure=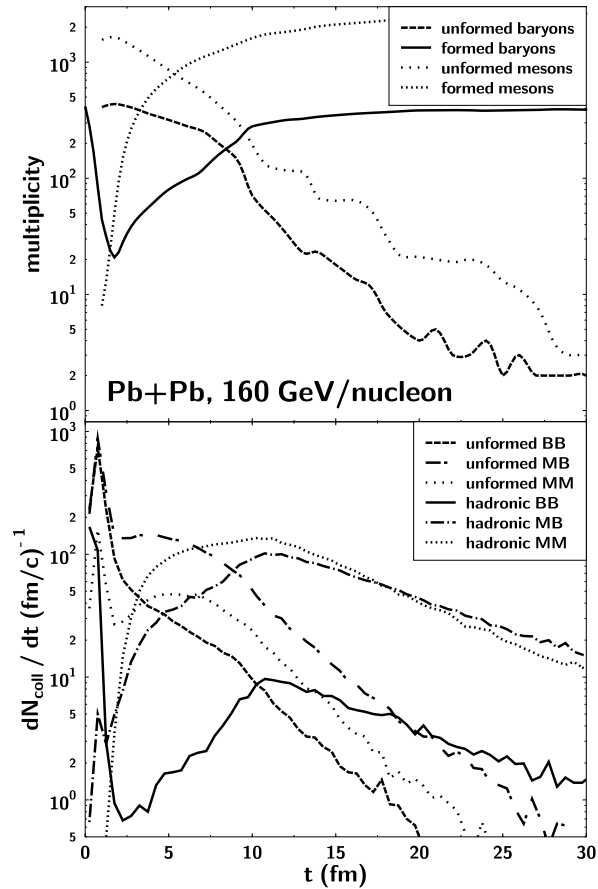,width=3.3in}}
\caption{\label{npart}
	Top: Time evolution of the multiplicity of hadrons and partonic
	constituents, divided into baryonic and mesonic contributions.
	Bottom: Collision rates for baryon-baryon (BB) and meson-meson (MM)
	collisions. The rates have been decomposed into interactions
	involving formed hadrons and those involving partonic constituents.}
\end{figure}

The time evolution of partonic constituents and  hadrons 
is shown in the upper frame of 
figure \ref{npart}. The first 5 fm/c of the 
reaction are dominated by the partonic constituents. 
The long-dashed and the dotted curves show the 
number of baryons and mesons contained in those constituents. 
In the case of leading-particles these can be interpreted as  
constituent (di)quarks or, for freshly born partons with small cross sections,
as excitation modes of the color field.

The lower frame of figure~\ref{npart}
shows the time evolution of the number of 
baryon-baryon (BB) and meson-meson (MM) collisions, both for ``hadronic''
and ``partonic'' interactions. ``Partonic'' interactions denote
interactions of {\em leading} (di)quarks either among themselves or with fully
formed hadrons. The early reaction stages, especially the
MM case, is clearly dominated by those ``partonic'' interactions. 
This number increases further if the scattering of the newly formed
partons is included.
Thus 
``partonic'' degrees of freedom significantly contribute both,
to the energy density, as well as to the collision dynamics in the
first 5 fm/c.

It should be noted that the "partonic"  collision
rates  can increase with the partonic cross section
during formation time: In this analysis all interactions during
formation time have been considered purely "partonic". Other
scenarios, however, include a "hadronic" contribution 
to the cross section which increases continuously
during $\tau_D$ and reaches its full hadronic value at the end
of $\tau_D$ \cite{gerland98a}.

\begin{figure}[tb]
\centerline{\psfig{figure=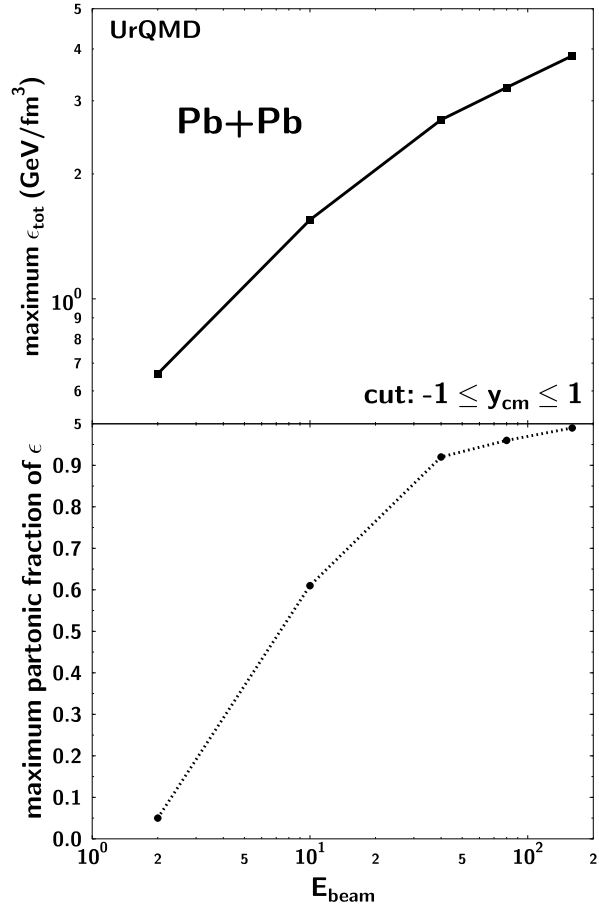,width=3.3in}}
\caption{\label{exfun}
	Top: excitation function of the maximum total energy density
	mid-rapidity hadrons experience. 
	Bottom: excitation function of the maximum ``partonic'' fraction
	of energy density. Already at a beam energy of 40 GeV/nucleon
	more than 90\% of the energy density is contained in partonic
	degrees of freedom at one time during the collision.
	}
\end{figure}

Do ``partonic''
degrees of freedom play any role at 10 GeV/nucleon, i.e. at the AGS?
The upper frame of figure~\ref{exfun} 
shows the maximum total energy density obtained in
central collisions of heavy nuclei as a function of incident beam energy,
starting from 2 GeV/nucleon and going  up to 200 GeV/nucleon. 
The energy density is obtained by the same method as used figure~\ref{edens}.
However, here ``partonic'' and ``hadronic'' contributions have been summed.
$\epsilon$ increases monotonously with the beam energy, 
reaching 
values as high as 4 GeV/fm$^3$ for SPS energies, which would seem
unreasonably high, if a purely hadronic 
scenario were used. 

The lower frame of figure~\ref{exfun} shows the maximum 
fraction of the energy density which is
contained in ``partonic'' degrees of freedom. Even at AGS,
energies already more than half of the energy density is due to
such ``partonic'' degrees of freedom, even though these do not yet dominate
the ``hadronic'' contributions.
At 40 GeV/nucleon, the maximum of the fraction of ``partonic'' energy
density is already $>90$\% of the total $\epsilon$.

The monotonous increase of the energy density 
excitation function does not imply that the excitation function of the
space-time volume of high {\em baryon density} shows the same behavior.
At AGS energies, $E_{lab} \sim 10$ GeV/nucleon, 
baryons still dominate the composition of the hadronic matter,
whereas at CERN/SPS energies, 200 GeV/nucleon, 
mesons constitute the largest fraction
of the hadronic matter. The maximum space-time volume of dense {\em baryonic}
matter can be reached at beam energies around 40 GeV/nucleon. A
detailed analysis of that regime, also with respect to experimental
signatures,  is presently underway \cite{weber98b}.


\begin{figure}[tb]
\centerline{\psfig{figure=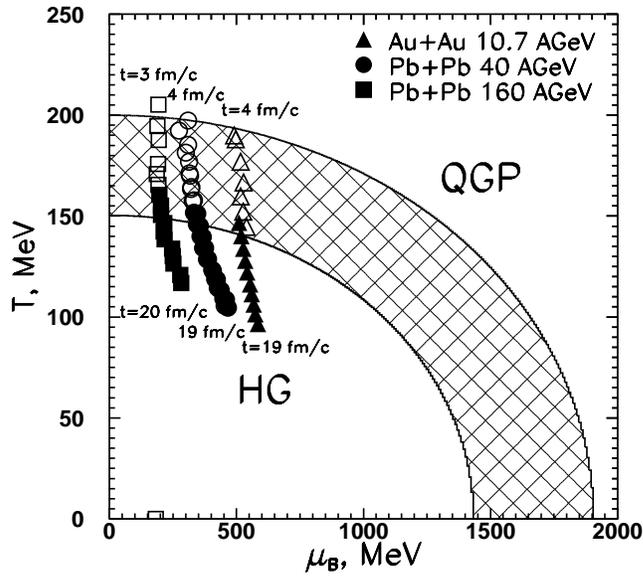,width=3.3in}}
\caption{\label{larisa1} Temperature versus baryonic chemical potential
predicted by the statistical model for the time 
evolution of the hadronic characteristics $\{\varepsilon,\ \rho_B,
\ \rho_S\}$ obtained within UrQMD in the central cell 
($2\times 2 \times 1$ fm$^3$ cell) of central 
A+A collisions at 10.7, 40 and 160 GeV/nucleon.
the open symbols show the non-equilibrium stage of the reaction, whereas
the full symbols denote the phase of local thermal equilibrium in
the central cell.
The two solid lines correspond to the phase-boundary between a
confined and deconfined phase  
calculated for two different bag constants, $B^{1/4}=227$ and 302 MeV
corresponding to $T_{\rm c}$=150 and 200 MeV at $\mu_B=0$.
 }
\end{figure}

The time evolution of temperature $T$ versus baryonic chemical potential
$\mu_B$ for Au+Au reactions at 10.6 GeV/nucleon and Pb+Pb reactions at 40 and
160 GeV/nucleon, respectively, is plotted in figure~\ref{larisa1}. 
The thermodynamic quantities $T$ and $\mu_B$ have been extracted by
fitting a statistical model to the quantities 
energy-, baryon- and strangeness density $\{\varepsilon,\ \rho_B,
\ \rho_S\}$ obtained from UrQMD in the central cell 
($2\times 2 \times 1$ fm$^3$ cell) of the heavy-ion reaction \cite{bravina98a}.
Here, the densities are calculated by
summing over all relevant partonic and hadronic degrees of freedom in the cell.
We see that the average $\mu_B$ in the reaction drops drastically 
with the initial collision energy, 
while the maximal temperature is growing and practically reaches 
the upper phase transition boundary with the critical 
temperature of $T_c$=200~MeV, 
as calculated with the MIT bag model (details of the used bag model 
can be found in \cite{goren97}).
However, during the early reaction stages matter in the central reaction
cell is neither fully hadronic, nor thermally and chemically equilibrated. 
A detailed analysis of velocity distributions and particle spectra 
in the central cell \cite{bravina98a} reveals that 
at approximately $t=2$~fm/c the velocity distributions of nucleons 
become isotropic in the central cell.
Pions, however, kinetically equilibrate much later, at $t \cong 8$ fm/c. This
effect is caused by the non-zero formation time for non-leading
particles. Full local thermal equilibrium (LTE) in the central cell 
(i.e. consistency of the particle spectra and yields with 
$T$ and $\mu_B$ extracted from $\{\varepsilon,\ \rho_B,\ \rho_S\}$) is 
first reached at $t \cong 10$ fm/c. 
To distinguish the later fully equilibrated
phase from the earlier reaction stage in which only nucleons show kinetic
equilibrium, the statistical model fits to the early reaction stage are
denoted by open symbols whereas the fits during the LTE phase are shown
in full symbols. During the phase of LTE
energy densities of $\sim 350$ MeV/fm$^{3}$ are never exceeded, 
i.e. a density above which the notion of separated hadrons 
would lose its meaning. 
However, the
main thermodynamic characteristics of the cell, $T$ and $\mu_B$
change rapidly with time.
This clearly demonstrates that a fireball type description
of hadronic matter is inadequate.

\begin{figure}[tb]
\centerline{\psfig{figure=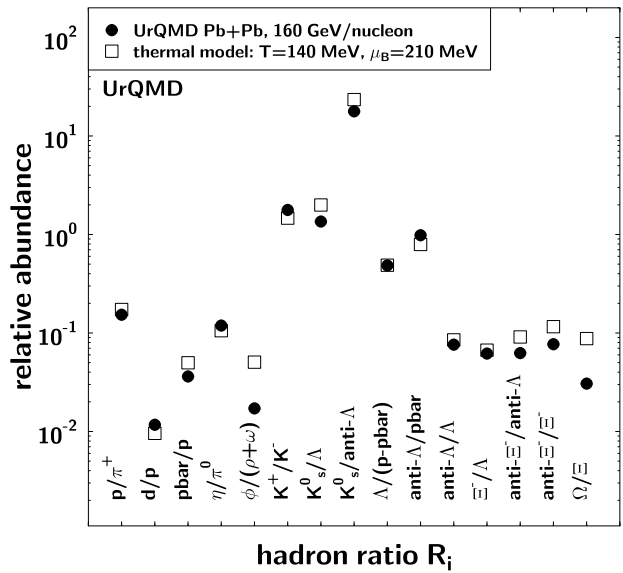,width=3.5in}}
\caption{\label{pbtherm} 
UrQMD prediction for hadron ratios in Pb+Pb
collisions at midrapidity (full circles). The open squares denote
a thermal model fit to the UrQMD calculation. The fit yields a 
temperature of T=140 MeV and a chemical potential of $\mu_B=210$ MeV.
}
\end{figure}
Let us now turn to the hadronic final state of the heavy-ion reaction.
Figure~\ref{pbtherm} shows a UrQMD prediction for hadron ratios in central
Pb+Pb reactions at 160 GeV/nucleon around mid-rapidity (full circles). 
The UrQMD prediction has been fitted with a statistical model, yielding
a temperature of T=140 MeV and a chemical potential of $\mu_B=210$ MeV.
However, this fit has been applied to the final hadron ratios after 
freeze-out. Here, the underlying assupmtion of the statistical model -- namely
a state of (global) equilibrium -- is not anymore valid, since the
break-up of the system and its freeze-out is governed by differences
in the interaction properties (i.e. cross sections) of the individual
hadron species.

\begin{figure}[tb]
\centerline{\psfig{figure=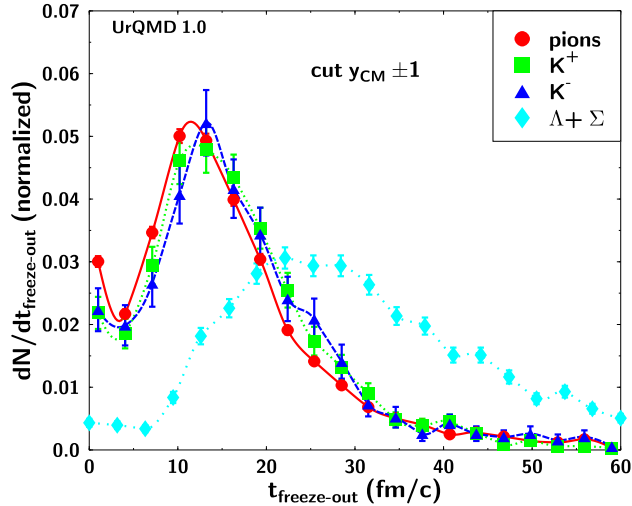,width=3.5in}}
\caption{\label{dndtf_pbpb} 
Normalized freeze-out time distribution 
for pions, kaons, antikaons and hyperons.
As with the freeze-out radii, the times for the meson species are very
similar. The hyperons again show a different behavior.
}
\end{figure}

To study the breakup of the system in greater detail,
let us turn to freeze-out distributions for individual hadron species:
Figure~\ref{dndtf_pbpb} shows the freeze-out time distribution 
for pions, kaons, antikaons and hyperons
at mid-rapidity in central Pb+Pb reactions at 160 GeV/nucleon.
The distributions have been normalized in order to compare the shapes
and not the absolute values.
In contrast to the situation at 2 GeV/nucleon, 
where each  meson species exhibits distinctly different
freeze-out time distributions \cite{bass98a}, 
all meson species here show
surprisingly similar freeze-out behavior -- 
the freeze-out time distributions all closely resemble each other.
Only the hyperons show an entirely different freeze-out behavior -- the
situation is even more extreme in the case of the multi-strange $\Omega$,
which exhibits a very sharp freeze-out time distribution,
distinctly different from all other hadron species \cite{bravina98a}. 
Whereas the common freeze-out characteristics of the mesons seem to 
hint at least at a partial thermalization, 
the hyperons show that even at SPS energies
there exists no common global freeze-out for all hadron species.
The same observation applies also to the distribution of 
transverse freeze-out radii.
Since these distributions have a large width, the average
freeze-out radius clearly does not define a freeze-out volume and
therefore estimates of the reaction volume or energy density based
on average freeze-out radii have to be regarded with great scepticism.
The large width of the freeze-out distributions is supported experimentally
by HBT source analysis which indicate the emitting pion source
to be ``transparent'', emitting pions from everywhere rather than from a
thin surface layer \cite{roland_heinz_qm97}.

\begin{figure}[tb]
\centerline{\psfig{figure=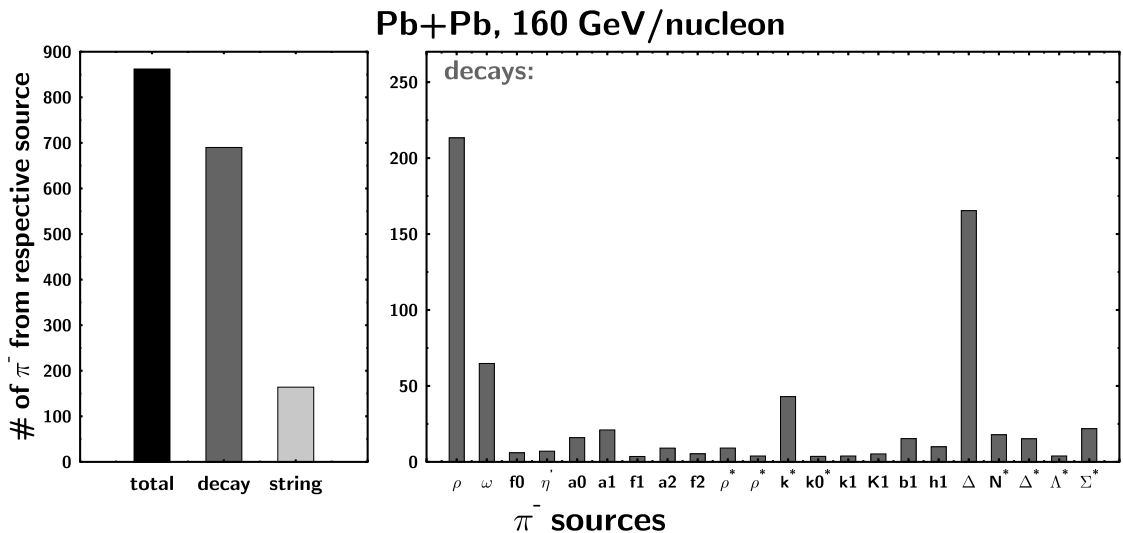,width=6.0in}}
\caption{\label{pi_origin} Pion sources in central Pb+Pb collisions at
CERN energies: 80\% of the final pions stem from resonance decays 
and 20\% from direct production via string fragmentation. Decay-pions
predominantly are emitted from the $\rho$ and $\omega$ mesons
and the $\Delta_{1232}$ resonance. }
\end{figure}

Figure~\ref{pi_origin} displays the respective sources from which 
negatively charged pions freeze-out. 
Only inelastic processes have been taken into account.
Approximately 80\% of the pions stem from resonance decays, only about
20\% originate from direct production via string fragmentation.
Elastic meson-meson or meson-baryon scattering adds a background of 
20\% to those numbers, i.e. 20\% of all pions scatter elastically
after their last inelastic interaction before freeze-out.
The decay contribution is dominanted by the $\rho, \omega$ and $k^*$ 
meson-resonances and the $\Delta_{1232}$ baryon-resonance -- no
weak decays have been taken into account in this analysis. 
However, more than 25\% of the decay-pions originate from a multitude
of different meson- and baryon-resonance states, some of which are shown
on the l.h.s. of figure~\ref{pi_origin}; e. g.
the two contributions marked $\rho^*$ 
stem from the $\rho_{1435}$ and the $\rho_{1700}$, respectively.

The analysis of the pion sources is of great importance for the 
understanding of the reaction dynamics and for the interpretation of
HBT correlation analysis results. The 20\% contribution of pions
originating from string fragmentation is clearly non-thermal, since
string excitation is only prevalent in the most violent, early 
reaction stages.

In summary, we have studied the evolution of relativistic Pb+Pb
reactions at CERN/SPS energies from the early non-equilibrium phase
through a stage of local thermal equilibration (in the central reaction cell)
up to its final hadronic freeze-out.
The importance of ``partonic'' degrees of freedom
in the early reaction stage
does not imply that an equilibrated
Quark-Gluon-Plasma has been formed. In the UrQMD approach the ``partonic''
phase has been modeled as an incoherent superposition of non-interacting
partonic constituents. 
Furthermore, these ``partons'' retain their original correlation into
hadrons -- deconfinement is not implemented into the present UrQMD approach.
The leading (di)quark interactions (among each
other and with fully formed hadrons) constitute 
an interacting ``mixed phase'' (for the constituent parton dynamics in 
this model, see, however \cite{gerland98a,spieles97c}).
In contrast, parton cascades \cite{geiger92a,geiger97a} 
allow for interactions among the partons only, while 
hadronic final state interactions are to a large extent neglected.

In the intermediate reaction phase, matter in the central cell can be 
viewed as hadrochemically equilibrated and exhibits an isentropic expansion.
However, this equilibrium stage is limited only to the central reaction
cell and breaks up in the late, dilute reaction phase close to freeze-out.
The freeze-out of the system, which is governed by the individual hadron
properties, has been studied in terms of freeze-out
radii for different hadron species and a source analysis for the contributions
of different microscopic processes to the final pion yield. A complex
freeze-out scenario emerges with species- and momentum dependent
broad freeze-out radius and time distributions.

S.A.B. acknowledges many helpful discussions with Berndt M\"uller.
This work has been supported by GSI, BMBF, Graduiertenkolleg ``Theoretische
und experimentelle Schwerionenphysik'', the Alexander v. Humboldt Foundation
and  DFG and DOE grant DE-FG02-96ER40945.

\end{document}